\documentclass[aps,prd,showpacs,floatfix,preprint]{revtex4}
\usepackage{amsmath,bm}
\usepackage{graphicx}
\usepackage{epsfig}
\begin{document}
\title{Melting of antikaon condensate in protoneutron stars} 
\author{Sarmistha Banik, Rana Nandi and Debades Bandyopadhyay} 

\affiliation{Astroparticle Physics and Cosmology Division and Centre for Astroparticle Physics, 
Saha Institute of Nuclear Physics, 1/AF Bidhannagar, Kolkata 700 064, India}

\begin{abstract}
We study the melting of a $K^-$ condensate in hot and neutrino-trapped 
protoneutron stars. 
In this connection, we adopt relativistic field theoretical models to describe
the hadronic and condensed phases. It is observed that the critical temperature
of antikaon condensation is enhanced as baryon density increases. For a
fixed baryon density, the critical temperature of antikaon condensation in a
protoneutron star is smaller than that of a neutron star. We also exhibit the 
phase diagram of a protoneutron star with a $K^-$ condensate.

\pacs{26.60.+c, 21.65.+f, 97.60.Jd, 95.30.Cq}
\end{abstract}

\maketitle

\section{Introduction}
Recent observation of a 2$M_{\odot}$ neutron star puts stringent conditions on
the equation of state (EoS) of dense matter in neutron star interior 
\cite{Demo}. It has been long conjectured that neutron star interior might
contain exotic phases of dense matter such as hyperons, 
Bose-Einstein condensates of antikaons and quarks. In certain cases, exotic
components of matter make the EoS softer resulting in maximum masses of neutron
stars below 2$M_{\odot}$. For example, the equations of state involving 
hyperons lead to maximum neutron star mass below 2$M_{\odot}$ when 
the hyperon-scalar meson coupling
is determined from existing hypernuclei data and hyperon-vector meson couplings
are estimated from SU(6) quark model \cite{mish,cbb}. Recently, it has been
demonstrated that the neutron star mass of 2$M_{\odot}$ could be achieved by
making hyperon-vector meson coupling stronger \cite{dcj1,dcj2,schm}. On the 
other hand, it was shown that the EoS including antikaon condensates might 
result in 2$M_{\odot}$ neutron stars if the magnitude of the attractive 
antikaon optical potential depth is 140 MeV or less \cite{pal}.   

The idea of antikaon condensation
in dense baryonic matter formed in heavy ion collisions as well as in neutron 
stars, started with the seminal work of Kaplan and Nelson \cite{Kap}. They 
investigated the problem within the $SU(3)_L\times SU(3)_R$ chiral perturbation
theory. Later detailed studies on antikaon condensation 
in neutron star interior were carried out in the chiral 
perturbation theory \cite{Bro92,Tho,Ell,Lee,Pra97} as well as meson exchange 
models \cite{mish,pal,Gle99,Mut,Kno,Li,Bani1,Bani2,Bani3,Bani4}. The first order
phase transition from nuclear matter to antikaon condensed matter was either
studied using Maxwell construction or Gibbs' phase equilibrium rules. It was 
observed that the threshold density for $K^-$ condensation was sensitive to 
the nuclear EoS as well as the strength of the attractive 
antikaon optical potential depth. 

Here we are interested in the melting of antikaon condensate in newly born
hot and neutrino-trapped protoneutron stars. The study of antikaon condensation
continued at finite temperatures in connection with the metastability
of protoneutron stars \cite{Pons} as well as the dynamical
evolution of the condensation \cite{Muto}. The critical temperature
of antikaon condensation in hot neutron stars, just after emission of trapped
neutrinos, was investigated for the first time \cite{Bani5}. We have recently 
investigated
the thermal nucleation of droplets of antikaon condensed matter in neutron 
stars after neutrinos were emitted as well as in protoneutron stars
\cite{Bani6,Bani7}. Further we made use of the results of our earlier 
calculation on the critical temperature of antikaon condensation in the thermal 
nucleation of antikaon droplets in neutron stars \cite{Bani5,Bani6}. However, 
there are no calculations on the critical temperature of antikaon condensation in
hot and neutrino-trapped protoneutron stars. This might have important 
implications to understand whether the droplet of antikaon condensed phase
would melt or not during the thermal nucleation in protoneutron stars.    
This motivates us to investigate the critical temperature of $K^-$ 
condensation in protoneutron stars. 

The organisation of the paper is the following. We discuss the model to compute
the composition and EoS involving $K^-$ condensate at finite temperature in 
Section II. Results are discussed in Section III. Section IV gives a summary.

\section{Formalism}
We investigate the melting of a $K^-$ condensate in hot and neutrino trapped
protoneutron stars. We adopt the finite temperature calculations of 
Ref.\cite{Pons,Bani5} to study this problem. The antikaon condensation is 
treated as a
second order phase transition from hadronic to $K^-$ condensed matter in 
protoneutron stars. Both phases of matter are made of
neutrons ($n$), protons ($p$), electrons, neutrinos, thermal $K^-$ mesons;
$K^-$ condensate only in the latter phase. Both phases maintain  local charge 
neutrality and  beta-equilibrium conditions. We can write down the total 
thermodynamical potential of both phases as
\begin{equation}
\Omega_{tot} = \Omega_N + \Omega_K + \Omega_L~.
\end{equation}  

We describe the hadronic phase using a relativistic field theoretical model
where baryons are interacting by the exchange of $\sigma$,
$\omega$ and $\rho$ mesons. The corresponding Lagrangian density  
is given by \cite{wal,bog},  
\begin{eqnarray}
{\cal L}_B &=& \sum_{B=n,p} \bar\Psi_{B}\left(i\gamma_\mu{\partial^\mu} - m_B
+ g_{\sigma B} \sigma - g_{\omega B} \gamma_\mu \omega^\mu
- g_{\rho B}
\gamma_\mu{\mbox{\boldmath t}}_B \cdot
{\mbox{\boldmath $\rho$}}^\mu \right)\Psi_B\nonumber\\
&& + \frac{1}{2}\left( \partial_\mu \sigma\partial^\mu \sigma
- m_\sigma^2 \sigma^2\right) - U(\sigma) \nonumber\\
&& -\frac{1}{4} \omega_{\mu\nu}\omega^{\mu\nu}
+\frac{1}{2}m_\omega^2 \omega_\mu \omega^\mu
- \frac{1}{4}{\mbox {\boldmath $\rho$}}_{\mu\nu} \cdot
{\mbox {\boldmath $\rho$}}^{\mu\nu}
+ \frac{1}{2}m_\rho^2 {\mbox {\boldmath $\rho$}}_\mu \cdot
{\mbox {\boldmath $\rho$}}^\mu ~,
\end{eqnarray}
where $\psi_B$ denotes the Dirac bispinor for baryons $B$, $m_B$ is the vacuum
mass and the isospin operator is ${\mbox {\boldmath t}}_B$. The scalar
self-interaction term \cite{bog} is
\begin{equation}
U(\sigma) = \frac{1}{3} g_2 \sigma^3 + \frac{1}{4} g_3 \sigma^4 ~.
\end{equation}

The thermodynamic potential per unit volume of the hadronic phase is given by 
\cite{Ser},
\begin{eqnarray}
\frac{\Omega_N}{V} &=& \frac{1}{2}m_\sigma^2 \sigma^2
+ \frac{1}{3} g_2 \sigma^3 + \frac{1}{4} g_3 \sigma^4  
- \frac{1}{2} m_\omega^2 \omega_0^2 
- \frac{1}{2} m_\rho^2 \rho_{03}^2  \nonumber \\
&& - 2T \sum_{i=n,p} \int \frac{d^3 k}{(2\pi)^3} 
[ln(1 + e^{-\beta(E^* - \nu_i)}) +
ln(1 + e^{-\beta(E^* + \nu_i)})] ~,  
\end{eqnarray}
where the temperature is related to $\beta = 1/T$, 
$E^* = \sqrt{(k^2 + m_N^{*2})}$ and the effective baryon mass 
$m_N^*=m_N - g_{\sigma N}\sigma$.
Neutron and proton chemical potentials are obtained from
$\mu_{n} = \nu_n + g_{\omega N}
\omega_0 - \frac {1}{2} g_{\rho N} \rho_{03}$ and
$\mu_{p} = \nu_p + g_{\omega N}
\omega_0 + \frac {1}{2} g_{\rho N} \rho_{03}$.

We can immediately calculate the thermodynamic quantities of the hadronic
phase such as the pressure $P_N = - {\Omega_N}/V$ and the energy density 
\begin{eqnarray}
\epsilon_N &=& \frac{1}{2}m_\sigma^2 \sigma^2
+ \frac{1}{3} g_2 \sigma^3 + \frac{1}{4} g_3 \sigma^4  
+ \frac{1}{2} m_\omega^2 \omega_0^2 
+ \frac{1}{2} m_\rho^2 \rho_{03}^2  \nonumber \\
&& + 2 \sum_{i=n,p} \int \frac{d^3 k}{(2\pi)^3} E^* 
\left({\frac{1}{e^{\beta(E^*-\nu_i)} 
+ 1}} + {\frac{1}{e^{\beta(E^*+\nu_i)} + 1}}\right)~.  
\end{eqnarray}
Similarly we can compute neutron and proton number densities which include 
contributions from both particle and antiparticles \cite{Pons,Bani5}.

Next we describe the antikaon condensed phase. The thermodynamic potential 
of this phase is calculated adopting the finite temperature 
treatment of antikaon condensation by Pons et
al. \cite{Pons}. In this case, we treat the (anti)kaon-baryon interaction in 
the same footing as the baryon-baryon interaction. 
The Lagrangian density for (anti)kaons in the minimal coupling scheme is 
\cite{Gle99,Bani2},
\begin{equation}
{\cal L}_K = D^*_\mu{\bar K} D^\mu K - m_K^{* 2} {\bar K} K ~,
\end{equation}
where the covariant derivative is
$D_\mu = \partial_\mu + ig_{\omega K}{\omega_\mu} 
+ i g_{\rho K}
{\mbox{\boldmath t}}_K \cdot {\mbox{\boldmath $\rho$}}_\mu$ and
the effective mass of (anti)kaons is
$m_K^* = m_K - g_{\sigma K} \sigma$.

The thermodynamic potential per unit volume in the antikaon condensed phase is
\cite{Pons}
\begin{equation}
\frac {\Omega_K}{V} = T \int \frac{d^3p}{(2\pi)^3} [ ln(1 - 
e^{-\beta(\omega_{K^-} - \mu)}) + 
 ln(1 - e^{-\beta(\omega_{K^+} + \mu)})]~,
\end{equation}
where the in-medium energies of $K^{\pm}$ mesons are given by
\begin{equation}
\omega_{K^{\pm}} =  \sqrt {(p^2 + m_K^{*2})} \pm \left( g_{\omega K} \omega_0
+ \frac {1}{2} g_{\rho K} \rho_{03} \right)~,
\end{equation}
and the chemical potential of $K^-$ mesons is $\mu = \mu_n -\mu_p$. The 
threshold condition for $s$-wave $K^-$ condensation is given by
$\mu = \omega_{K^{-}} =   m_K^* - g_{\omega K} \omega_0
- \frac {1}{2} g_{\rho K} \rho_{03}$~.

The number density of (anti)kaons is given by
\begin{equation}
n_K = n_K^C + n_K^{T}~,
\end{equation}
which has contributions from the condensate $n_K^C$ and thermal (anti)kaons 
given by,
\begin{equation}
n_K^{T} =  
\int \frac{d^3 p}{(2\pi)^3} 
\left({\frac{1}{e^{\beta(\omega_{K^-}-\mu)} 
- 1}} - {\frac{1}{e^{\beta(\omega_{K^+}+\mu)} - 1}}\right)~.  
\end{equation}

In the antikaon condensed phase, only thermal (anti)kaons contribute to 
the pressure $P_K = -{\Omega_K}/{V}$; the condensate does not contribute
to the pressure.
The energy density of (anti)kaons is given by 
\begin{equation}
\epsilon_K = m_K^* n_K^C + \left( g_{\omega K} \omega_0
+ \frac {1}{2} g_{\rho K} \rho_{03} \right) n_K^T
+
\int \frac{d^3 p}{(2\pi)^3} 
\left({\frac{\omega_{K^-}}{e^{\beta(\omega_{K^-}-\mu)} 
- 1}} + {\frac{\omega_{K^+}}{e^{\beta(\omega_{K^+}+\mu)} - 1}}\right)~.  
\end{equation}

Finally, we can calculate thermodynamic quantities of 
electrons, neutrinos and their antiparticles from the thermodynamic potential
per unit volume
\begin{equation}
\frac{\Omega_L}{V} = - T \sum_l g_l \int \frac{d^3 k}{(2\pi)^3} [ ln(1 + 
e^{-\beta(E_l - \mu_l)})
+ ln(1 + e^{-\beta(E_l + \mu_l)})]~,
\end{equation}
where $g_l$=2 for electrons and 1 for neutrinos.

We use mean field approximations in this calculation. Mean meson fields are
obtained by minimising the total thermodynamic potential of Eq. (1).
The total energy density in the condensed phase is 
$\epsilon = \epsilon_N + \epsilon_K + \epsilon_L$. Similarly the total entropy
per baryon is given by 
$S = ({\cal S}_N + {\cal S}_K + {\cal S}_L)/n_b$, where ${{\cal S}_N}$,
${\cal S}_K$ and ${\cal S}_L$ are entropy densities of the hadronic phase,
antikaon condensed phase and  
leptons, respectively \cite{Pons,Bani5}. 

Further the EoS of hot and lepton-trapped matter is constrained by the charge 
neutrality and $\beta$ equilibrium conditions which are given by,
\begin{eqnarray}
n_p - n_K - n_e=0~,\\
\mu = \mu_n - \mu_p = \mu_e - \mu_{\nu_e}~.
\end{eqnarray}

\section{Results and Discussion}
We use the GM1 parameter set for nucleon-meson coupling constants which
reproduce the nuclear matter saturation properties i.e. binding energy -16 MeV, 
saturation density ($n_0$) 0.153 $fm^{-3}$, asymmetry energy coefficient 32.5 
MeV, effective nucleon mass ($m_N^*/m_n$) 0.70 and incompressibility 
$K = 300$ MeV \cite{Bani2,Gle91}. 

Kaon-vector meson couplings are obtained from the quark model
and isospin counting rules \cite{Gle99,Bani5}. Further the kaon-scalar meson 
coupling is estimated from the real part of antikaon optical potential depth at
normal nuclear matter density \cite{pal,Gle99,Bani2}. It was already noted that 
the value of antikaon optical potential depth varied widely from  
$U_K = -180 \pm 20$ MeV as obtained from the analysis of kaonic atom 
data \cite{Fri94,Fri99,Gal07} to $\sim -60$ MeV in the chiral model \cite{Tol}. 
Earlier we observed that the EoS involving the $K^-$ condensate became very soft
when the antikaon potential was highly attractive. Such an EoS 
resulted in maximum neutron star mass below 2$M_{\odot}$ \cite{pal}. After the
discovery of 2$M_{\odot}$ neutron star, the EoS including exotic matter 
is severely constrained. Therefore we perform this calculation for 
$U_K = -120$ MeV which results in a maximum neutron star mass of 
2.08$M_{\odot}$ \cite{pal}. Kaon-scalar coupling constants are taken from 
Table II of Ref. \cite{Bani2}. 

We consider a set of values for entropy per baryon $S$ in this 
calculation. Further we take lepton fraction $Y_L = Y_e + Y_{\nu_e} = 0.35$ in 
the protoneutron star. These correspond to several snapshots in the evolution
of neutrino-trapped and hot protoneutron stars. For a fixed entropy per baryon,
the temperature varies from the center to the surface in the protoneutron star.
This is demonstrated in Figure 1. We highlight this for entropy per baryon
$S=2$. The temperature increases with baryon density in this case.
Here, we also show the corresponding scenario in a hot and neutrino-free 
neutron star for $S=2$. Higher temperature is obtained in the neutron star.
Further we note that as soon as the antikaon condensate appears in the
protoneutron star, there is a drop in the temperature compared with the case 
without the condensate. 

Populations of different particle species in the $\beta$-equilibrated 
protoneutron star matter are shown with baryon density for $S=2$ case in
Figure 2. We observe that the matter is populated with thermal kaons well 
before the onset of the $K^-$ condensate. This also leads to the enhancement of
the proton fraction. The threshold density of $K^-$ condensation in the
hot and neutrino-trapped protoneutron star matter for entropy per baryon $S=2$ 
is 4.2$n_0$. On other hand, $K^-$ condensation in the neutrino-trapped matter 
at zero temperature sets in at 3.68$n_0$. Similarly, threshold densities of 
antikaon condensation in neutrino-free neutron stars with $U_{K}=-120$ MeV are 
3.16$n_0$ for $S=2$ and 3.05$n_0$ for $S=0$. The role of finite temperature is 
to shift the threshold of antikaon condensation to higher density. 
After the appearance of the condensate, $K^-$ density 
($n_K^C$) increases rapidly resulting in higher proton number 
density in the protoneutron star as evident from Fig. 2.

We discuss the thermal effects on the EoS and protoneutron star masses. 
Pressure versus energy density is plotted for entropy per baryon $S=2$ and
$S=0$ in Figure 3. We do not find any appreciable change in the EoS due to
thermal effects. This result has important significance for the calculation of 
thermal nucleation of the antikaon condensed phase in protoneutron stars and 
justifies the use of a zero temperature EoS in that case \cite{Bani7}. Maximum 
protoneutron star masses for $S=2$ and $S=0$ cases are 2.228 and 
2.214$M_{\odot}$, respectively. The corresponding central density for $S=2$ 
case is 5.3$n_0$ and it is 5.59$n_0$ for $S=0$.  

Next we investigate the melting of the condensate in protoneutron star matter
as the interior temperature increases. As we
follow the evolution of a protoneutron star through several snapshots, we 
consider different values of entropy per baryon. We demonstrate the variation  
of entropy per baryon with temperature for several fixed values of
baryon densities and $U_K = -120$ MeV in Figure 4. 
The condensate density 
($n_K^C$) is a function of both baryon density and temperature. The condensate
disappears above a certain temperature known as critical temperature ($T_C$). 
To obtain this critical 
temperature, we show the ratio of the condensate density ($n_K^C(T)$) at finite
temperature to that ($n_K^C(T=0)$) of zero temperature as a function of 
temperature for several fixed baryon densities and $U_K = -120$ MeV in 
Figure 5. In each case corresponding to a fixed baryon density, 
the condensate density diminishes with increasing temperature resulting ultimately in the
meltdown of the condensate at the critical 
temperature of the condensation. We find that the critical temperature 
increases as baryon density increases. 

In an earlier calculation, we estimated the critical 
temperature of $K^-$ condensation for hot and neutrino-free neutrons stars
\cite{Bani5}. When we compare the critical temperatures in both cases for a
certain baryon density for example 4.4$n_0$, the critical temperature in the 
protoneutron star has a smaller value than that of the neutron star as
evident from Figure 6. This may be attributed to more heat content in the 
protoneutron star than that of the neutron star. For example, at 
temperature $T=40$ MeV, entropy per baryon corresponding to $Y_L=0.35$ and 
$Y_{\nu_e}=0$ are 1.6 and 1.3, respectively. This finding might be very useful
in understanding the thermal nucleation of droplets of antikaon condensed matter
in protoneutron stars \cite{Bani7}.

As we know the critical temperature as a function of baryon density, 
we can construct a phase diagram of protoneutron star matter involving a $K^-$
condensate. The phase diagram as temperature versus baryon density 
is shown in Figure 7. The solid line representing critical temperatures 
separates the condensed phase (shaded region) from the hadronic phase. 

\section{Summary}
We have investigated the critical temperature of $K^-$ condensation in 
$\beta$ equilibrated hot and neutrino-trapped protoneutron stars for antikaon 
optical potential depth $U_K = -120$ MeV within the framework of field 
theoretical models at finite temperature. The critical temperature of antikaon 
condensation in hot and neutrino-trapped protoneutron stars increases with 
baryon density. It is also noted that the critical temperature of antikaon
condensation in a protoneutron star is smaller than that of a hot and
neutrino-free neutron star. This result might play an important role
in the thermal nucleation of droplets of antikaon condensed 
matter in protoneutron stars. 


\newpage 
\vspace{-2cm}

{\centerline{
\epsfxsize=12cm
\epsfysize=14cm
\epsffile{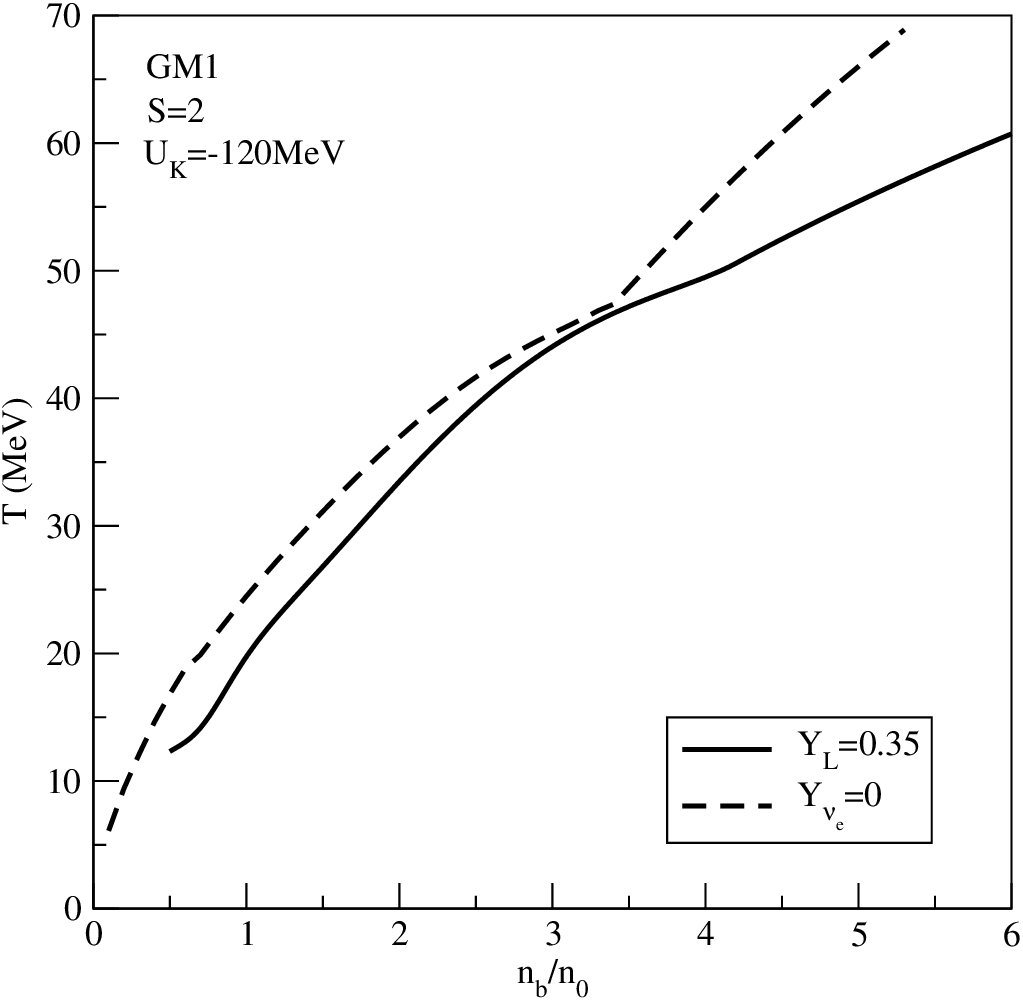}
}}

\vspace{4.0cm}

\noindent{\small{
FIG. 1. Temperature is plotted with normalised baryon density for entropy
per baryon $S = 2$, $Y_L=0.35$ and antikaon optical potential depth at 
normal nuclear matter density $U_{K} = -120$ MeV.}}

\newpage
\vspace{-2cm}

{\centerline{
\epsfxsize=12cm
\epsfysize=14cm
\epsffile{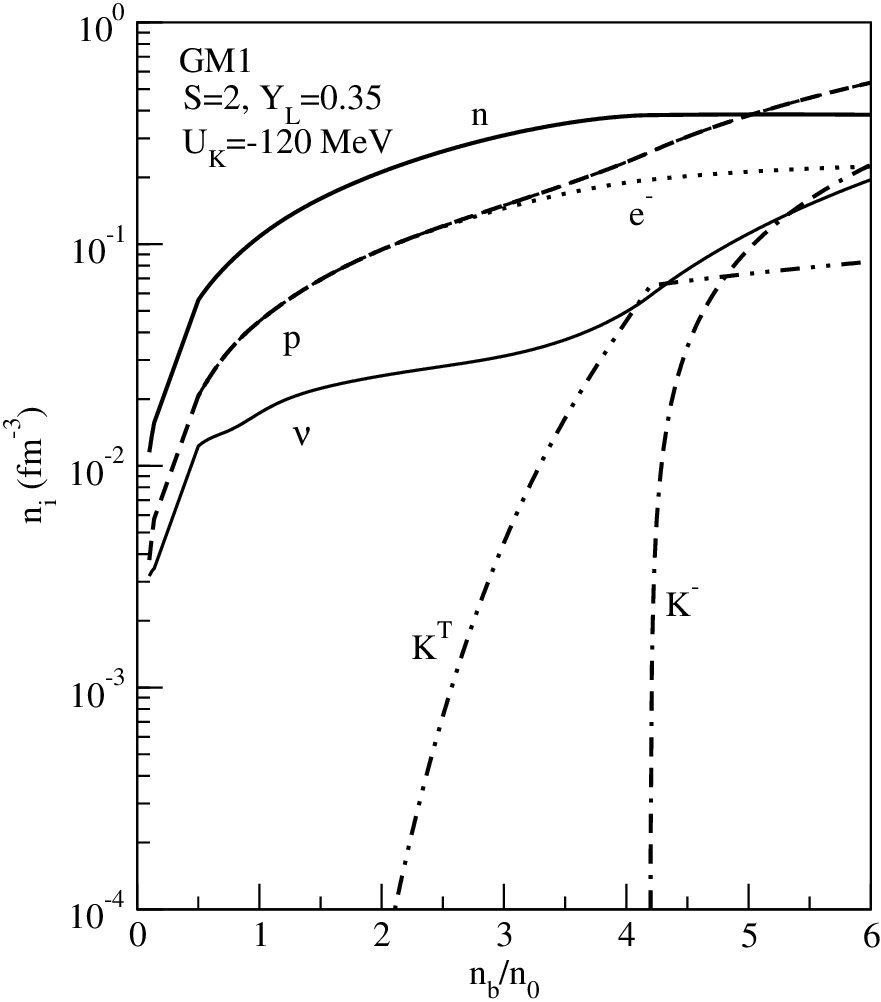}
}}

\vspace{4.0cm}

\noindent{\small{
FIG. 2. Populations of different particle species in 
$\beta$-equilibrated hot and lepton-trapped matter including a $K^-$ 
condensate are shown as a function of normalised baryon density for entropy
per baryon $S = 2$, $Y_L=0.35$ and antikaon optical potential depth at 
normal nuclear matter density $U_{K} = -120$ MeV.}}
\newpage
\vspace{-2cm}

{\centerline{
\epsfxsize=12cm
\epsfysize=14cm
\epsffile{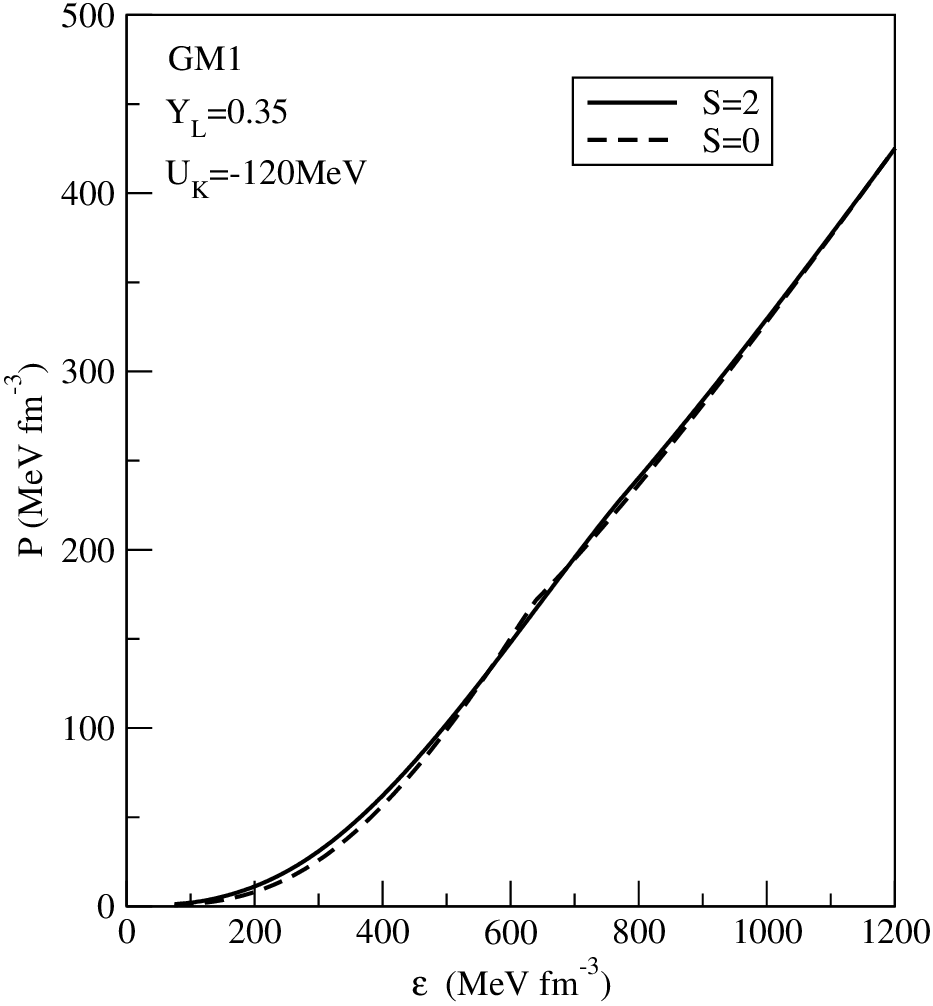}
}}

\vspace{4.0cm}

\noindent{\small{FIG. 3. Pressure is plotted with energy density for 
entropy per baryon $S=2$ and $S=0$ and $Y_L=0.35$ and antikaon optical 
potential depth at normal nuclear matter density $U_{K} = -120$ MeV.}}

\newpage
\vspace{-2cm}

{\centerline{
\epsfxsize=12cm
\epsfysize=14cm
\epsffile{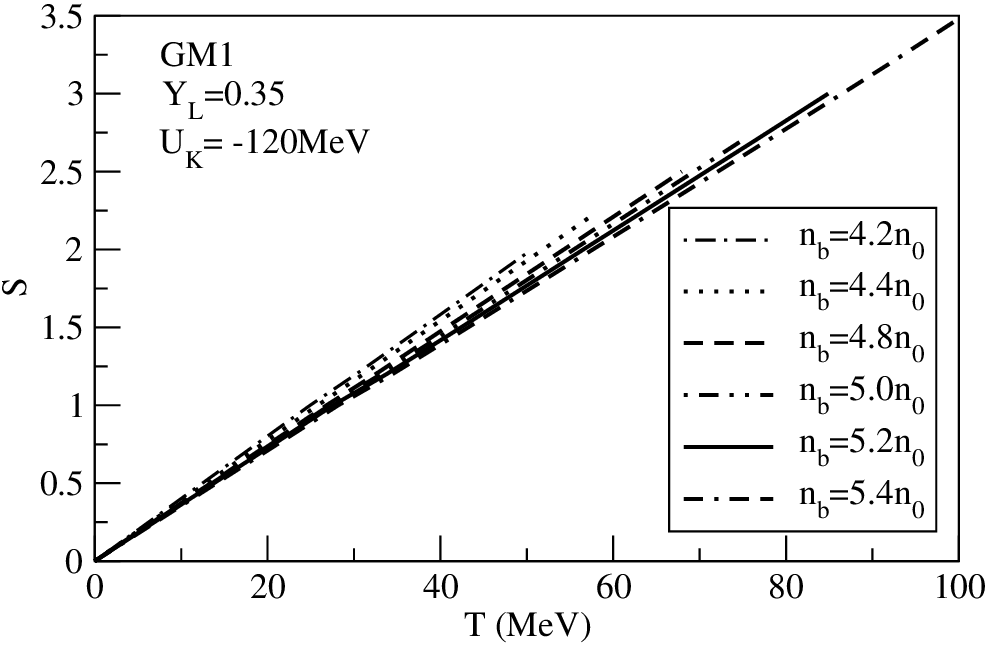}
}}

\vspace{4.0cm}

\noindent{\small{
FIG. 4. Entropy per baryon is plotted with temperature for 
fixed baryon number densities, $Y_L=0.35$ and antikaon optical potential 
depth at normal nuclear matter density $U_{K} = -120$ MeV.}}

\newpage
\vspace{-2cm}

{\centerline{
\epsfxsize=12cm
\epsfysize=14cm
\epsffile{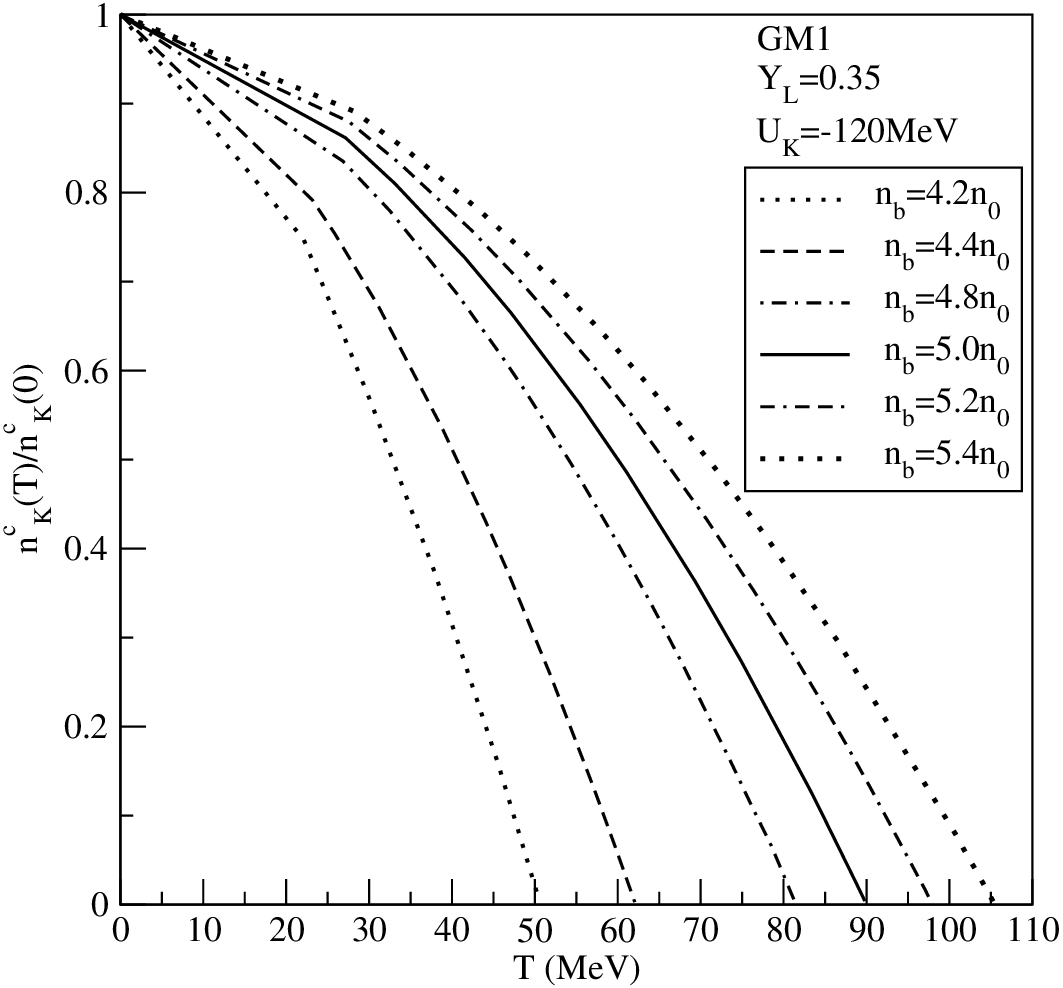}
}}

\vspace{4.0cm}

\noindent{\small{
FIG. 5. The ratio of the condensate density at a nonzero
temperature to that of zero temperature is plotted with temperature for 
fixed baryon number densities, $Y_L=0.35$ and antikaon optical potential 
depth at normal nuclear matter density $U_{K} = -120$ MeV.}}

\newpage
\vspace{-2cm}

{\centerline{
\epsfxsize=12cm
\epsfysize=14cm
\epsffile{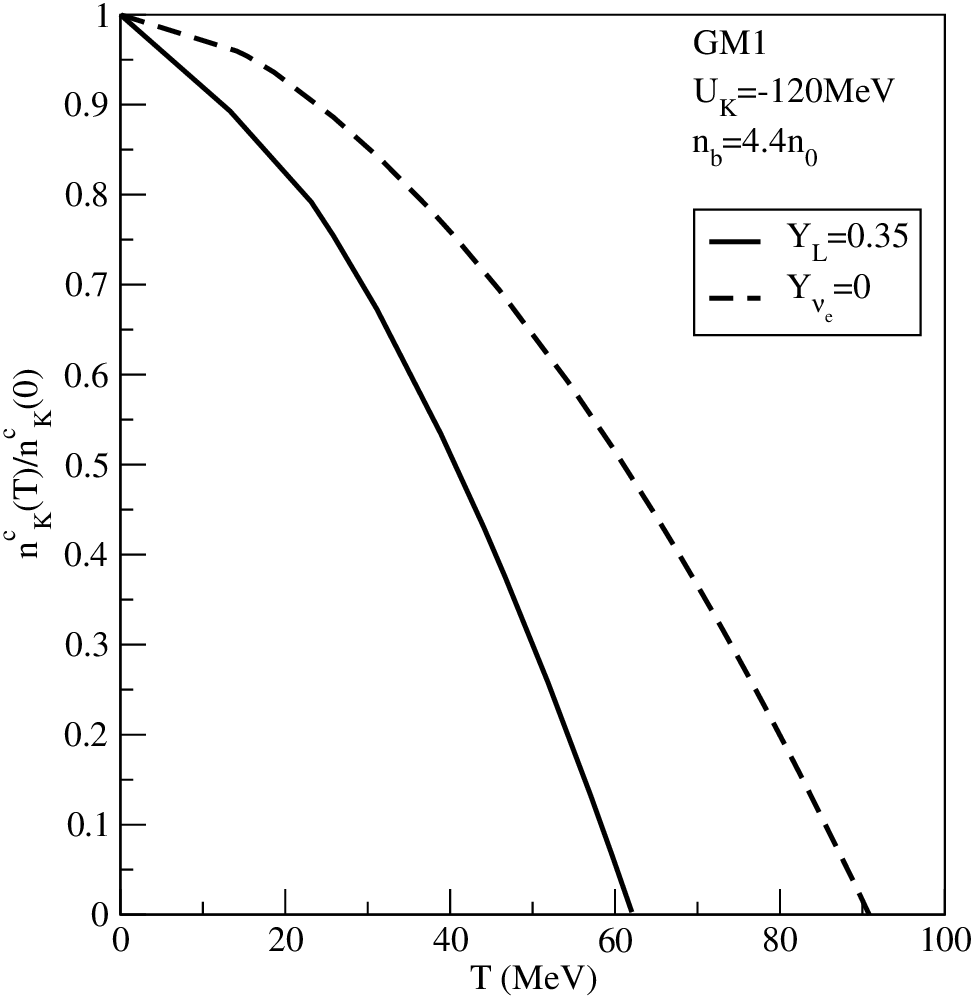}
}}

\vspace{4.0cm}

\noindent{\small{
FIG. 6. The ratio of the condensate density at a nonzero
temperature to that of zero temperature is plotted with temperature for a 
hot and neutrino-trapped protoneutron star as well as a hot and neutrino-free
neutron star at a fixed baryon number density 4.4$n_0$ and 
antikaon optical potential depth at normal nuclear matter density 
$U_{K} = -120$ MeV.}}

\newpage
\vspace{-2cm}

{\centerline{
\epsfxsize=12cm
\epsfysize=14cm
\epsffile{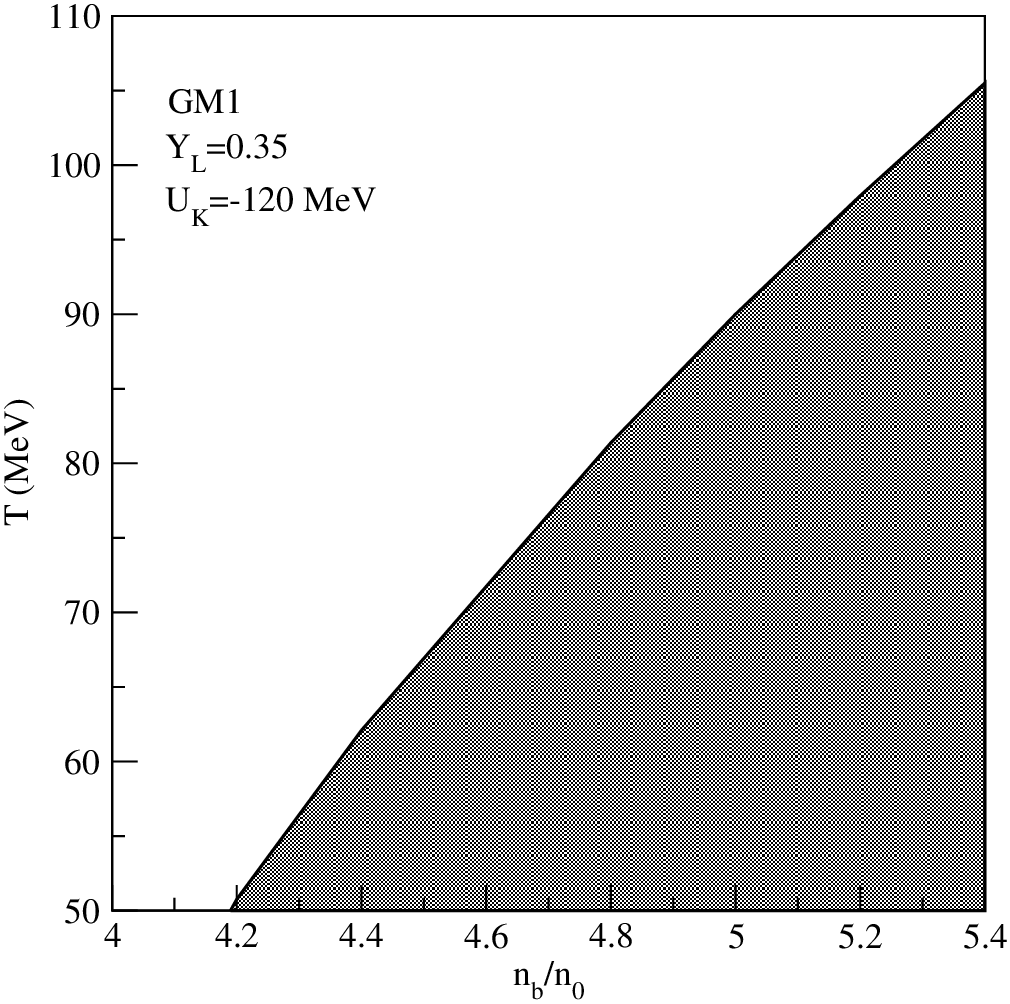}
}}

\vspace{4.0cm}

\noindent{\small{
FIG. 7. Phase diagram of protoneutron star matter with $K^-$ condensate. The 
solid line corresponds to critical temperatures of $K^-$ condensation for 
$Y_L=0.35$ and antikaon optical potential depth at normal nuclear matter 
density $U_{K} = -120$ MeV.}}
\end{document}